IAA-CU-15-11-52

# IITMSAT Communications System – A LeanSat Design Approach

*Akshay K. Gulati\*, Shubham Chavan\*\*, Joseph Samuel, Sampoornam Srinivasan, Pradeep Shekhar, Akshat Dave, Aditya Sant, Sourbh Bhadane, Mayug Maniparambil, Vishnu Prasad S, Dhanalakshmi Durairaj, David Koilpillai, Harishankar Ramachandran.*

## Abstract

IITMSAT is a student-built nanosatellite mission of Indian Institute of Technology Madras, Chennai, India. The objective is to study the precipitation of high energy electrons and protons from Van-Allen radiation belts to lower altitude of 600-900 km due to resonance interaction with low frequency EM waves. The unique communications system design of IITMSAT evolves from the challenging downlink data requirement of 1 MB per day in the UHF band posed by the mission and the satellite's payload, SPEED (Space based Proton and Electron Energy Detector).

---
\* Indian Institute of Technology Madras, India, akshaygulati26@gmail.com
\*\* Indian Institute of Technology Madras, India, shubhamc.iitm@gmail.com

To ensure continuous downlink data stream in the short LEO passes, a robust physical layer protocol was designed to counter time-varying aspects of a Space-Earth telecom link. COTS integrated radios were found to be incompatible for both on-board and ground segments, since they had constraints on either the frame structure, on the modulation scheme, or bandwidth. Hence each of the segment's system architecture was carefully divided into smaller blocks to a level that can make use of COTS modules. To ensure flexibility for customization, the Ground Station configuration was built on a SDR framework. As all the selected GS hardware was COTS, the focus was only on software and hardware tests. For the on-board communications system, two types of design alternatives exist for each module. The first option is a custom design wherein a module is developed from scratch using discrete components. The selected components should be at least industrial, automotive or extended temperature grade, and designed for high-vibration environments. The other option is an integrated design wherein an electronics COTS module can be directly plugged into the subsystem. This module is evaluated by carrying out vibration and thermal tests. If an integrated module is low-cost and meets the design requirements, it is preferred over a custom design. This design criteria was used as a reference to ensure minimum acceptable reliability. All the modules were individually verified against their specifications. In order to carry out performance tests under simulated link conditions, an RF attenuation test setup was designed that can work at extreme temperatures. Burn-In tests for 72 hours at ambient and extreme temperatures were carried out. Integrated tests indicate all IITMSAT design requirements have been met. Hence a robust communications system has been validated. The time taken for development of on-board telecom and GS was less than a year and was achieved at a low cost which agrees to a LeanSat approach.

## Introduction

Since the inception of IITMSAT project, the aim was to focus on developing a scientifically useful mission even though this is the first satellite from IIT Madras. Since an innovative and fairly complex payload was to be built in-house i.e. in a university framework, the spacecraft bus had to be designed in a simple, risk free and low-cost manner. With this philosophy in mind, the design on-board and ground station communication systems have been developed that satisfy the project requirements. The details of the IITMSAT payload, SPEED and a description of science is mentioned in [1].

## Communication system architecture

The IITMSAT communication system comprises of three links, namely downlink, uplink and beacon.
The average downlink data requirements were fixed to 1 MB per day. If it is assumed that there is continuous downlink data stream in the short LEO passes over a single ground station, a raw data rate of 9.6 kbps will be sufficient to achieve complete data downlink. A robust physical layer protocol was designed to counter time-varying aspects of a Space-Earth telecom link. Since the protocol implements an error correction algorithm, the data rate doubles to 19.2 kbps. The details of this protocol are given in [2]. This fits within the bandwidth constraints posed by UHF amateur satellite radio link. According to a NASA JPL study [3], GMSK comes to 2nd position in terms of bandwidth efficiency after FQPSK-B. Implementing GMSK is much easier due to the availability of a large number of power efficient GSM based transceiver chips.
The uplink requirements are not stringent. To ensure a high link budget, a low data rate of 1000 bps was decided upon. FSK was chosen to ensure simplicity.
Beacon is designed to be a reliable transmitter system that downlinks basic housekeeping data and can work even if the main transmitter or receiver fails. Beacon transmits in OOK modulation scheme at a low data rate of for short durations of 4 seconds at regular intervals of 30 seconds and a low power of 20 dBm.

## 1. On-board system

A few key system level decisions were made to reduce the risk of the onboard communication system. The architecture is illustrated in Fig 1.

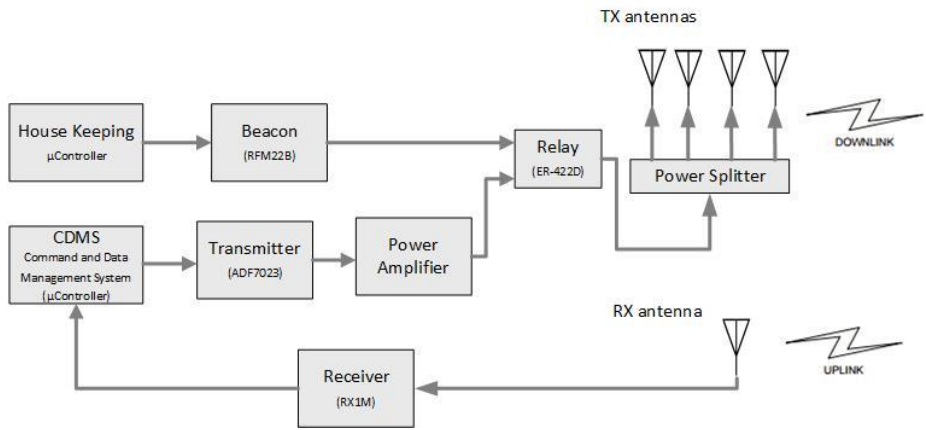

Fig. 1. IITMSAT on-board communication architecture.

### Antennas

The requirement on IITMSAT is of a non-deployable antenna along with an omnidirectional radiation pattern. This puts limitations on all COTS antennas. Fixed monopole antennas typically need a large reflective ground plane and there is no available satellite face for this. The only feasible option is to use Inverted F antennas (IFA).

For the receiver, a single long IFA antenna is enough. However for the transmitter, four small IFAs are required to provide the omnidirectional pattern. This implies that the transmitter RF output needs to be equally divided and fed into the four antennas. This can be achieved by using a 4-way 0º power divider. Since both the transmitter and the beacon have to use the same antenna system, an RF switch has been implemented that ensures only one RF output is connected to the input of power divider.

Radio

The transmitter and receiver modules have been made independent of each other. This is done to ensure minimum coupling between the transmitter and receiver and to ensure maximum flexibility in the selection of COTS modules.

## 2. Ground system

The ground segment architecture is based on a Software Defined Radio (SDR) platform. COTS transceivers could not be used in the proposed 20 kHz bandwidth downlink due to the fact that they possess a narrow IF filter of bandwidth only 5 to 15 kHz. SDR was chosen as the solution to overcome this problem. SDR also provides other benefits and are described towards the end of this paper. The ground system architecture is illustrated in Fig 2.

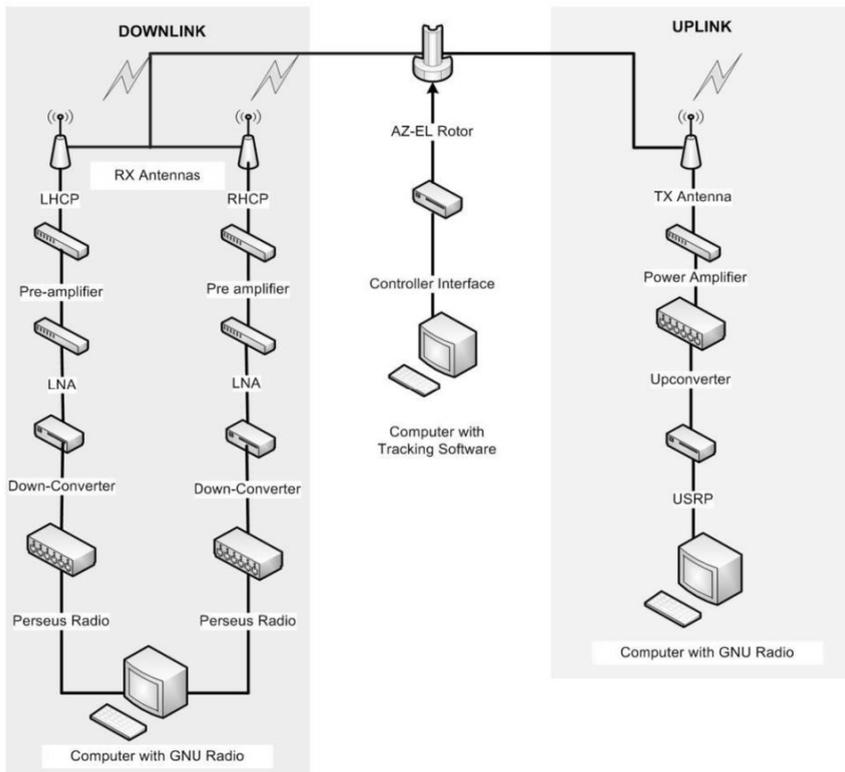

Fig. 2. IITMSAT ground system architecture.

The polarization of the downlink signals undergo a shift from Left Hand Circular Polarization (LHCP) to Right Hand Circular Polarization (RHCP) or vice versa during a pass. This is due to the asymmetric polarization in the radiation pattern of the Transmitter antennas. To combat this, there are two receiver antennas for downlink – one for LHCP and one for RHCP. Both antennas have identical RF chains connected to it.

## Hardware selection and tests

To reduce development time, the guiding philosophy has been to use COTS modules wherever possible and ensure requirements are met by carrying out rigorous integration tests.

### *1. On board system*

The onboard system comprises of the transmitter, receiver, beacon, RF switch and power divider.

Transmitter

There are many COTS transmitters available in market. However most were found to be incompatible since they had constraints on either the frame structure, on the modulation scheme, or data rate. A summary of this is given in Table 1.

| COTS Transmitter | Cost (USD) | Compatibility issues |
|---|---|---|
| Helium Radio | 6100 /- | Cannot implement custom frame format |
| XDL Micro | 480 /- | Cannot implement custom frame format |
| NanoCom U482C | 8500/- | Data rate and Modulation incompatible and cannot implement custom frame format |
| NanoCom AX100 | NA | Cannot implement custom frame format |
| Intrepid UHF Radio | 5000/- | Cannot implement custom frame format |

Table 1. List of COTS radios with details of their compatibility issues.

Though some of the product datasheets such as Helium radio and XDL Micro did mention that a transparent protocol is possible, however the customer support was not good and the student team were not able to make the radio work. This resulted in loss of money and time. No single transmitter was available in market that satisfies our requirements. To build from scratch using a transceiver chip seemed to be the only possible solution. Some of the team members attempted to make a custom transmitter. However a transmitter comprises of many elements and they found it hard to integrate them into a single board.

Instead of searching for a single PCB solution, it was decided to build it using smaller COTS plug-in modules. The transmitter could comprise of two separate modules, a low-power transmitter and a power amplifier. This helped the team expand their search and noticed that some of the evaluation boards made by chip manufacturers could be our solution. Analog Devices has many low power transceiver chips and one chip, ADF 7023 satisfies the design requirements well. The ADF7023 evaluation board consists of a mother board and a daughter board connected to each other by three board to board connectors. The ADF7023 module with its motherboard is illustrated in Fig 3. The features of ADF7023 are mentioned in Table 2.

|                | Features                  | IITMSAT Specifications |
|----------------|---------------------------|------------------------|
| Frequency band | 431- 464 MHz              | 437.505 MHz            |
| Data rate      | 1 kbps to 300 kbps        | 19.2 kbps              |
| Modulation     | FSK, GFSK, OOK, GMSK      | GMSK                   |
| RF output power| -20 to +13.5 dBm          | 12 dBm                 |
| Digital Interface | SPI                    | SPI                    |
| Protocol       | Custom, Hardware Defined  | Custom                 |

Table 2. Features of ADF7023 [4].

One quick and feasible solution was to use the ADF7023 daughter board and make our own motherboard and attach the same connectors that are placed on the ADF7023 motherboard. This task does not require any RF circuit design. However the vibration survivability of the board to board connectors is not guaranteed. To counter the effects of the vibration, appropriate adhesive at the interfaces and a cable tie based support system has been designed.

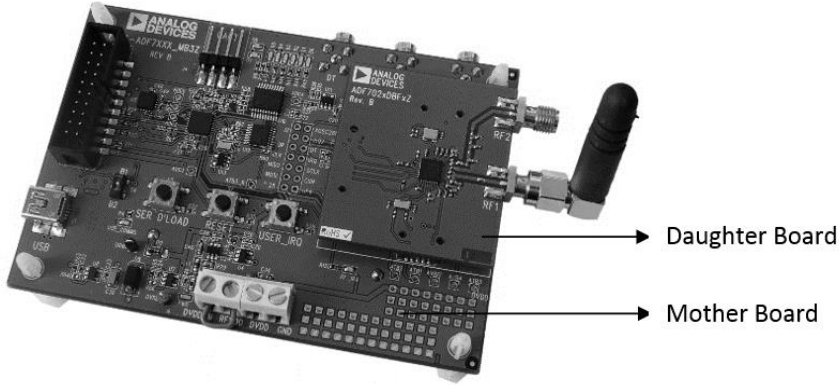

Fig. 3. ADF7023 Evaluation board. The daughter board is plugged into the mother board [5].

A similar approach was carried out for the power amplifier (PA). Though no plug-in module was found, two PA evaluation boards satisfy all performance requirements. They have an option of getting attached to a mother board via fasteners. For the electrical interface, both of these PAs have 2.54" inch berg strip connectors. A reliable solution is to attach wires between the mother board and the PA evaluation board. The details of the power amplifier boards are mentioned in Table 3. RF6886 is designed to deliver 36 dBm output at a good efficiency of 54 %, however we need to transmit at 30 dBm only. Input/output tests were carried out to optimize the efficiency for it. Though the power efficiency of ASX601 is much better, RF6886 requires a voltage that can be easily supplied by the on-board electrical power system without any change. To reduce the risk, it was decided to develop with both the options in parallel.

| Power amplifier evaluation board | PA efficiency | Power output | Optimized voltage Input |
|---|---|---|---|
| RF6886 | 33% | 30 dBm | 3.0 V |
| ASX601 | 68% | 30 dBm | 5 V |

Table 3. List of power amplifier options can be mounted via fasteners to a mother board.

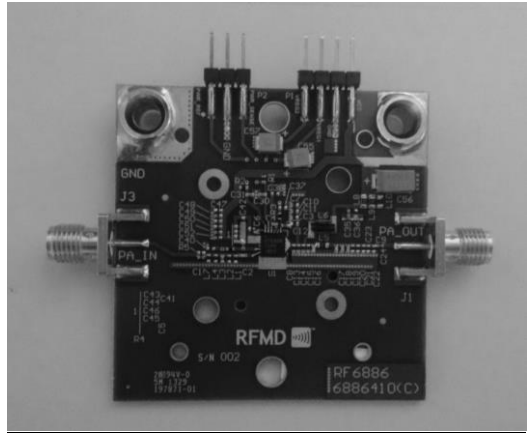

Fig. 4. RF6886 Evaluation board.

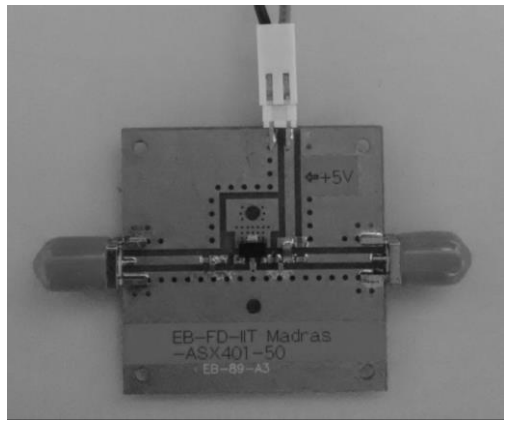

Fig. 5. ASX401 Evaluation board.

### Beacon

The requirements of the beacon are not stringent. It was preferred to have a single COTS module rather than having a separate transmitter module. HopeRF has a wide range of radio modules and many of them satisfy the requirements.

RFM22B from HopeRF has also flown on a few other satellite missions. The details of its space heritage are provided in Table 4.

Moreover a breakout board is available from SparkFun[1] and Modtronics[2] for easy prototyping. This helped the team in saving valuable time. RFM22B is a very versatile system. All major communication parameters such as modulation scheme, data rate, frequency, output power levels, packet format etc. are easily programmable. The features of RFM22B and IITMSAT specifications are mentioned in Table 5.

| Satellite | Launch Information | Current status |
|---|---|---|
| $50Sat | 21 Nov 2013 | Active |
| T-LogoQube | 21 Nov 2013 | Active |

Table 4. Space heritage of RFM22B [6][7][8]

| | Features | IITMSAT Specifications |
|---|---|---|
| Frequency band | 415 - 453 MHz (resolution of 156.25 Hz) | 437.505 MHz |
| Data rate | 0.125 - 40 kbps | 120, 600 bps |
| Modulation | FSK, GFSK, OOK, | OOK |
| RF output power | +1 to +20 dBm | 20 dBm |
| Digital Interface | SPI | SPI |
| Protocol | Custom, Hardware Defined | Custom |

Table 5. Features of RFM22B and IITMSAT specifications [9][10]

The IITMSAT beacon structure has two packets that are transmitted one after other. The short beacon and long beacon packet are transmitted at 40 bps and 120 bps for a duration of 3 seconds and 1 seconds respectively, every 30 seconds. The transmitted data includes the satellite's call sign and some important health parameters like the solar panels' temperature, battery status, voltages, current etc.

Initially the beacon hardware is initialized via SPI as the communication protocol and the registers corresponding to following parameters are set. The beacon data is dumped on the FIFO buffer of the Beacon hardware which can hold 64 bytes of data. Since the beacon's buffer size is not sufficient to hold the entire beacon data at a time, it is filled dynamically and the transmission is done till the entire message is transmitted. The beacon module is placed in standby mode at the end of the transmission.

---

[1] https://www.sparkfun.com/products/retired/9821
[2] http://modtronicsaustralia.com/shop/rfm22b-s2-rf-transceiver-breakout-board-v2/

Testing was done to verify the power consumption and power output of the beacon. A known random data packets were transmitted and the output was verified by receiving signal by the Ground station segment. The results are mentioned in Table 6.

| Specification | Expected value | Observed value |
| --- | --- | --- |
| Peak power consumption | 85 mA @ 3.3 V = 280.5 mW | 95 mA @ 3.3 V = 313.5 mW |
| Peak output power | 20 dBm | 21.2 dBm |
| Frequency | 435.505MHz | 434.503MHz |

Table 6. Test results of RFM22B [10]

To verify its robustness, performance tests were carried out at different temperatures from 10ºC to 95ºC. The module worked well and transmitted till 90ºC beyond which it failed to transmit.

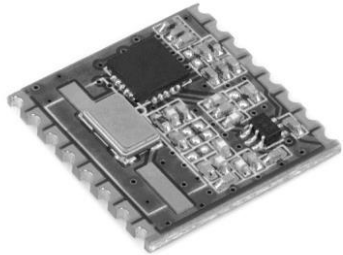

Fig. 6. RFM22B module.

RF switch

A RF relay is used to switch between the transmitted downlink signals of beacon and the transmitter. The output of the switch is connected to the input of the power divider. There are two type of RF switches, namely, relay based and semiconductor based. RF switch is a critical item as its failure will end the mission. Since semiconductors are susceptible to space radiation, a radiation resistant semiconductor based RF switch, PE9354 from Peregrine Semiconductor was identified. Similarly a vibration resistant Mil grade Relay, ER422D from Teledyne Relays was chosen. It has space heritage and is used in many ISRO satellites. The comparison of these two options are tabulated in Table 7. The performances are comparable, however the cost of PE9354 is much higher than ER422D. Hence it was decided to use ER422D as the RF switch.

| Characteristics | ER422D-5 | PE9354 |
|---|---|---|
| Insertion loss | 0.33dB | 0.45dB |
| Return loss | 23dB | 23dB |
| Isolation | -35 dB | -55dB |
| Power supply | 5V | 3V |
| Cost (in USD) | 93 /- | 1128 /- |

Table 7. Comparison chart of RF Switch options [11[12]

Teledyne's ER-422D is a magnetic latching relay and requires no holding power after the contacts have transferred. It also has an internal diode for coil transient suppression. The coils need to be energized using a 5V supply. The switching is controlled by a microcontroller. Hence an appropriate a transistor driver was designed using transistors. The transistor used is 2N3019, a military grade NPN leaded silicon transistor device. The design of the driver circuit is described in Fig. The rated minimum operate pulse is of 4.5ms width. As a safety measure, the designed pulse duration is 4-5 times this duration (25ms). The ER422D relay is much better in terms of insertion and return losses as compared to PE9354. The low isolation across contacts is not much of a concern as only one of the downlink transmitters will be sending the transmitted signal at one point of time.

These characteristics were verified by performing S-parameters tests using a low cost Vector Network Analyzer, VNWA 3 from SDR Kits at 435MHz.

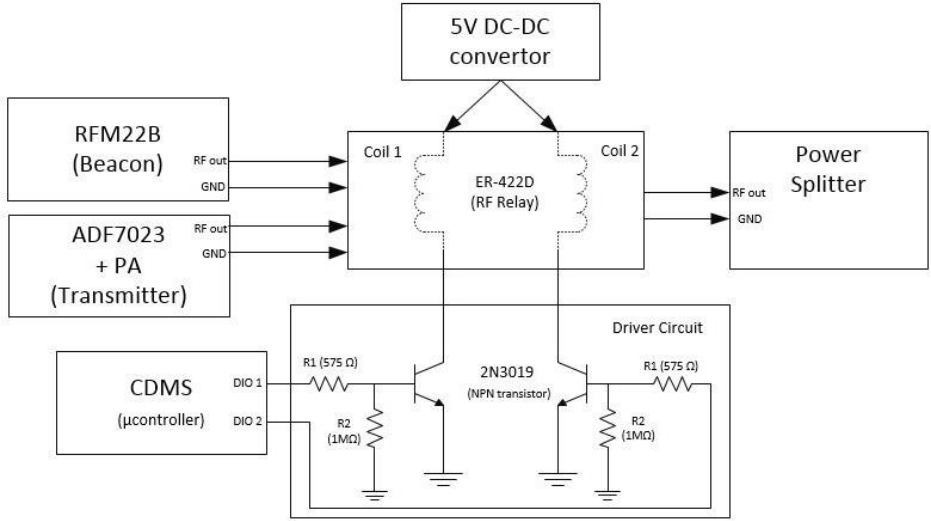

Fig. 7. Design of a RF Switch using a relay and a driver circuit.

Most of the electrical characteristics of the relay were also verified and are summarized in Table 8.

| Characteristics | Expected | Observed |
|---|---|---|
| Insertion loss | 0.33dB | 0.258dB |
| Return loss | 23dB | 22.68dB |
| Isolation | -35 dB | -30 dB |
| Operate time | 1.5 ms | 0.34 ms |
| Contact bounce | 3 ms | 1.8 ms |
| Contact resistance | 0.225 Ω | 0.184 Ω |
| Coil resistance | 61 Ω | 60.3 Ω |

Table 8. Performance Testings of RF Switch [11]

Hot-and-cold chamber testing was carried out to check the functionality at different temperatures. The set/reset coil voltage and coil current were measured and summarized in Table 9.

| Temperature (ºC) | Coil voltage (V) | Coil current (mA) |
|---|---|---|
| 9.5 | 1.832 | 0.034 |
| 20 | 2.0555 | 0.034 |
| 32 | 2.273 | 0.036 |
| 50 | 2.339 | 0.037 |
| 80 | 2.471 | 0.038 |
| 90 | 2.613 | 0.038 |

Table 9. Thermal tests of the RF switch

### COM TX

A custom transmitter mother board (COM TX) had to be designed that can support the ADF7023 & PA daughter boards. Beacon and RF switch are also a part of COM TX.

COM TX has a mechanically challenging design. The RF output of the ADF7023 is terminated to an SMA female connector. This has to be connected to the input of PA board with an SMA cable. Minimum bending radius of the SMA cable was calculated using a NASA handbook and has been adhered. Positions of the daughter boards has been accordingly adjusted.

COM TX has been functionally tested and satisfies all design requirements. The integrated COM TX PCB is shown in Fig. 8.

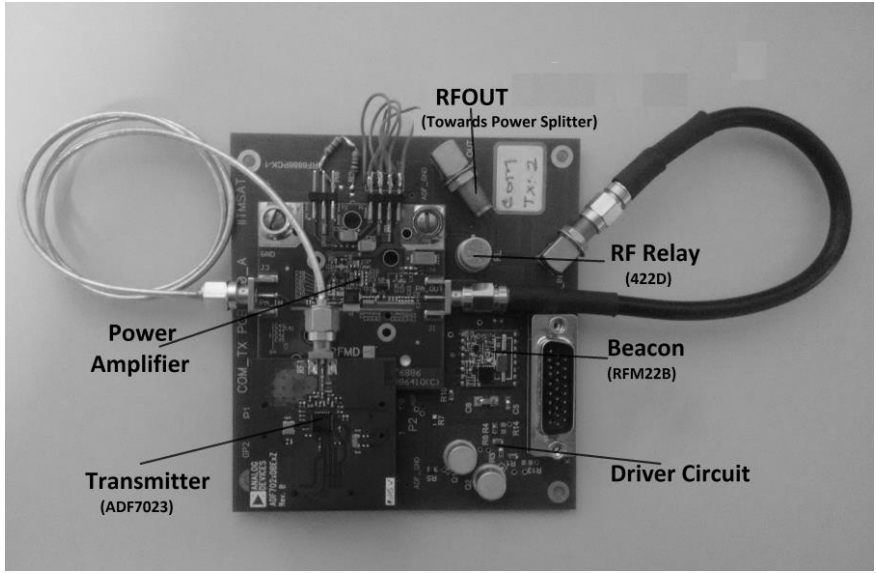

Fig. 8. COM TX PCB.

Receiver

There are no strict requirements on the frame structure of for up-link. However a custom packet protocol would help in improving the efficiency of data transfer since the data rate is low. Owing to a limited SPI interfaces in the on-board computer, a UART digital interface is preferred. RX1M from Radiometrix was selected and its key features are mentioned in Table 10.

| Specification | Expected value |
|---|---|
| Data rate | 1000 bps |
| Modulation | FSK |
| Sensitivity | -118dBm |
| Current | 13 mA |
| Voltage | 4.5 -15 V |
| Receiver blocking | -85dB |

Table 10. Specifications of RX1M [13]

The receiver uses a physical layer protocol that is very robust and rejects erroneous packets. This improves the system robustness. However the receiver hardware does not have any constraints on the packet structure. This has helped in implementing a custom packet protocol.

The receiver was tested from -5ºC to 50ºC. In this temperature range, the sensitivity was receiver found to be -106.8 dBm.

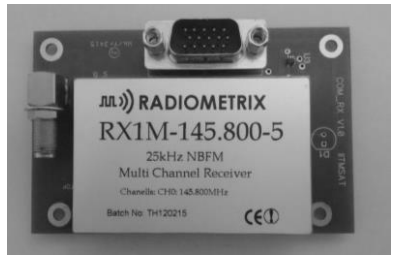

Fig. 9. COM RX PCB.

Power divider

To ease of routing of the SMA cables in the satellite the power divider has to be a separate unit that is placed on the spacecraft bottom panel. A power divider that has good RF characteristics and is robust will be a good choice. After a good survey, two such modules were identified, namely DL445 from TRM Microwave and PD-1000-4S from UMCC. A comparison chart of these two modules is given in Table 11. Most of the features are comparable. To have compact packaging and low mass, DL445 was chosen.

3rd IAA Conference On University Satellite Missions And Cubesat Workshop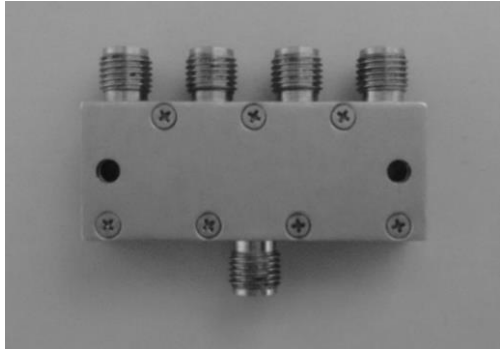
Fig. 10. DL445 Power divider.

IAA-CU-15-11-52

| Specification | DL445 | PD-1000-4S |
|---|---|---|
| Input power | 1.5 W | 4 W @ Load VSWR=∞ |
| VSWR IN | 1.16:1 (22.6 dB) | 35 dB |
| VSWR OUT | 1.19:1 (21.2 dB) | 24 dB |
| Phase Balance | 0.4° to 0.7° | ±0.7° |
| Isolation | 30 dB | 30 dB |
| Insertion loss | 0.4-0.6 dB | 0.5-0.6 dB |
| Mass | 19 g | 87.9 g |
| Temp op. (°C) | -40°C to +85°C | -55°C to +95°C |
| Temp Storage (°C) | -55°C to +125°C | -60°C to +110°C |
| Cost (for 3 nos.) | USD 805 each | USD 275 each |
| Vibration | MIL-STD-202 method 204, test condition A | MIL-STD-202F, Method 204D, Cond. B (.06" double amplitude, or 15G) |
| Shock | MIL-STD-202 method 213, test condition G | MIL-STD-202F, Method 213B, Cond. B (75G, 6mSec) |
| Thermal Shock | MIL-STD-202, method 107, test condition A | MIL-STD-202F, Method 107D, Cond. A (5 cycles) |

Table 11. Specifications of COTS power dividers [14][15]

### 2. Ground system

Most of the components in the ground system are COTS. The details are mentioned in Table 12.

| Component | Company | Part No. |
|---|---|---|
| RX Antenna | M2 | 436CP42UG |
| TX Antenna | M2 | 2MCP14 |
| Antenna Rotor | Alfa Radio | AlfaSpid RAS Az/El |
| Preamplifier | SSB | SP-7000 |
| Down-converter | SSB | DCM-70 |
| SDR (Downlink) | Microtelecom | Perseus SDR |
| Power Amplifier | Minicircuits | ZHL-100W-52+ |
| SDR (Uplink) | NI / Ettus Research | USRP N210 with WBX daughterboard |

Table 12. Product tree of ground system

Downlink

GNU Radio is an open source SDR platform and is widely used. Due to expertise in our institute, this was chosen to be the SDR platform for IITMSAT ground system. Perseus Radio offers very good sensitivity and is compatible with GNU radio. It was suitable for IITMSAT requirements. However it can process RF signals up to 32 MHz only. To use this receiver, an appropriate down-converter, DCM-70 was selected. The Mission Control System is based on a website based platform, Django and interfaces with GNU radio using python scripts.

Uplink:

The ground station transmitter has also been developed using the SDR framework. There was a proprietary physical layer protocol that was in-built in the on-board receiver hardware. This initially constrained the team to use the company's transmitter for uplink. However after analyzing the signals of the transmitter (TX1M) in GNU Radio and Matlab for various combinations, the team was able to decode the signals. Now GNU radio is able to transmit in the same way as TX1M. Standard SDR hardware that is compatible with GNU radio is USRP from Ettus Research[3].

Benefits gained using SDR

Even though filter issues with COTS radios was the primary reason for moving to SDR, it was realized subsequently that SDR offered huge benefits and advantages to a small satellite ground station setup. To the best of our knowledge, no other small satellite team has used SDR in their ground station. The use of SDR gave many advantages.

---

[3] http://www.ettus.com/

The main advantage of SDR is its flexibility. Changes in physical layer parameters like changing center frequency could be made at the click of a button.

The GUI interface given by GNU Radio (which is similar to LabVIEW) is extremely user friendly and any person with basic computer knowledge can operate the SDR-based ground station. Moreover, GNU Radio is open source and there is a big and active community available for help and support. Anyone with basic knowledge can create new blocks or modify existing ones in GNU Radio and hence modify the algorithms. This is in stark contrast to a pure hardware implementation in which changing some parameter is both time and resource consuming.

SDR also helps in Doppler integration. Doppler frequency offset prediction is done using GPredict software which runs on the same PC and is interfaced with GNU Radio through a TCP port. If hardware systems had been used, then we would have needed to dynamically feed the Doppler information generated by the prediction software (running on PC) to be hardware setup.

SDR also allows to share the decoded data with other computers through the Internet. SDR also enables smooth integration of physical layer data with other front-end user interface software (such as payload data analysis software) used in the ground station.

SDR gave two big benefits in designing the downlink receiver. The first is that it enabled us to design and implement custom physical layer protocols and algorithms. This was a great advantage given the high downlink data requirements of IITMSAT. If we had not used SDR, the only other option would have been to use commercial amateur radio receivers which employ protocols like AX.25. Such protocols have large overheads which would have made the downlink of large amounts of data highly challenging. SDR enabled us to marvelously overcome this challenge by allowing us to create a custom, low-overhead and robust physical layer protocol. The details of the physical layer design and algorithms is described in [2].

The second advantage that SDR in downlink receiver design was that it enabled us to test and prototype various receiver algorithms very easily. This was especially important while designing algorithms to handle the custom physical layer protocols. Algorithms could be developed in MATLAB and could be easily implemented and tested in GNU Radio without depending on any hardware components. Performance of the receiver could be easily analyzed using BER simulations from GNU Radio itself. These would have been extremely difficult if the team had resorted to a hardware based design.

SDR implementation allows us to easily store baseband samples in the PC. In addition to archiving, this gives another advantage. These samples can be processed later using complex and sophisticated algorithms (which could not be run real-time) and hence extract more meaningful data out of noisy signals.

In the case of uplink, SDR is being used to emulate TX1M. SDR allowed us to reverse engineer the physical layer protocol of TX1M and RX1M and implement it in our ground station.

The modular structure of an SDR implementation allows other satellites with similar requirements to reuse the developed design and code. A screenshot of the GNU radio program illustrates the same. Refer Fig 11. Thus, the authors strongly believe that the use of SDR can revolutionize the lean satellite ground station designs. Upcoming ground stations can switch to SDR from old amateur-radio-based hardware.

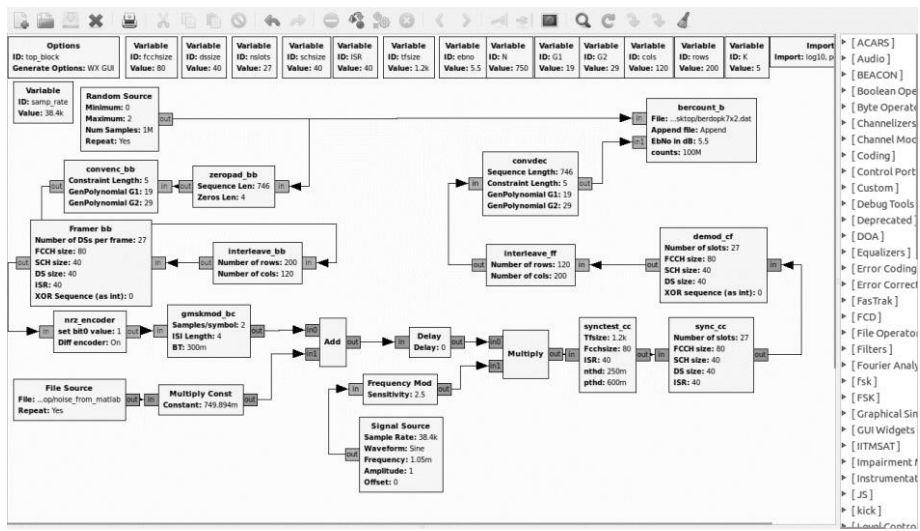

Fig. 11. Screenshot of GNU Radio

## Timeline

The development time of the on-board system has been less than a year. The detailed timeline is mentioned in Table 13. Component selection took a significant chunk of the time. Often not enough time is given to this initial stage. If adequate time is spent during component selection, cost can reduced and reliability can be increased.

| Component | Component Selection | Procurement | Prototyping | Fabrication | Tests | Total |
|---|---|---|---|---|---|---|
| Beacon | 2 | 0.5 | 3 | 1 | 0.5 | 7 |
| RF Switch | 4 | 0.5 | 2 | 1 | 0.5 | 8 |
| Transmitter | 6 | 0.5 | 3 | 1 | 0.5 | 11 |
| Power amplifier | 1 | 0.5 | 0.25 | 1 | 2 | 4.75 |
| Receiver | 2 | 1 | 0.5 | 1 | 2 | 6.5 |
| Power divider | 4 | 2 | 0.5 | - | - | 6.5 |

Table 13. Development time for IITMSAT COM on-board system (in months)

## Cost

The approach has been to make use of low cost COTS modules. The details of the cost budget for one COM TX and one COM RX are given in Table 14. The development costs are not added.

| Part | Cost (USD) |
|---|---|
| Beacon (RFM22) | 5.50 |
| Transmitter (ADF7023) | 93.75 |
| PA (RF6886) | 150.00 |
| Relay (ER442D-5) | 81.00 |
| Discrete parts (Resistors, capacitors and transistors) | 50.00 |
| SMA connectors (RA) 4 nos. | 120.00 |
| Connector (26 Pin HD D-Sub M) | 31.00 |
| Connector (15 Pin HD D-Sub M) | 19.00 |
| Receiver (RX1M) | 113.00 |
| Power Filter (BNX002) 2 nos. | 6.00 |
| Voltage Reference (LT1761) | 5.40 |
| PCB | 50.00 |
| SMA Cables | 100.00 |
| Total | 824.65 |

Table 14. Cost Budget of IITMSAT on-board COM system

## Conclusion

A low cost, reliable transmitter and receiver satellite radio has been developed using COTS modules for very unique non-standard requirements. The development time has been less than a year. Rigorous functional and performance tests have provide us an idea of the reliability of the design. A successful LeanSat design approach has been achieved.

## Acknowledgement

The authors deeply acknowledge the work of the following team members whose contribution was invaluable: Shreesha S, Mahesh K R, Kirubha T, Krishan P, Sukhdeep S K, Dheeraj M P, Nandu S, Rohit S, Vishal B M, Rangeeth R, Sanchit G, Shashank Y and Aman V. We thank Mr. Rajangam for the various technical inputs at various stages of the project. The authors acknowledge Dr. Krishna Chivukula, Indo-MIM Pvt. Ltd for the financial support to the project. The authors also acknowledge the support provided by scientists of ISRO Satellite Centre for the review of entire hardware and software at various stages of the project.